\documentclass{emulateapj}
\usepackage{color}
\usepackage{float}
\usepackage[normalem]{ulem}
\usepackage{hyperref}

\begin{document}

\title{The influence of metallicity on stellar differential rotation \\and magnetic activity}

\author{Christoffer Karoff}
\affil{Department of Geoscience, Aarhus University, H{\o}egh-Guldbergs Gade 2, 8000, Aarhus C, Denmark}
\affil{Stellar Astrophysics Centre, Department of Physics and Astronomy, Aarhus University, Ny Munkegade 120, 8000, Aarhus C, Denmark}
\email{karoff@phys.au.dk}

\author{Travis S. Metcalfe}
\affil{Space Science Institute, 4750 Walnut Street, Suite 205, Boulder CO 80301, USA}

\author{{\^A}ngela R. G. Santos}
\affil{Space Science Institute, 4750 Walnut Street, Suite 205, Boulder CO 80301, USA}
\affil{Instituto de Astrof{\' i}sica e Ci{\^e}ncias do Espa{\c c}o, Universidade do Porto, CAUP, Rua das Estrelas, PT4150-762 Porto, Portugal}
\affil{Departamento de F{\' i}sica e Astronomia, Faculdade de Ci\^{e}ncias, Universidade do Porto, Rua do Campo Alegre 687, PT4169-007 Porto, Portugal}
\affil{School of Physics and Astronomy, University of Birmingham, Birmingham B15 2TT, UK}

\author{Benjamin T. Montet}
\affil{The Department of Astronomy and Astrophysics, The University of Chicago, 5640 S. Ellis Ave, Chicago IL 60637, USA}

\author{Howard Isaacson}
\affil{Department of Astronomy, UC Berkeley, Berkeley, CA 94720, USA}

\author{Veronika Witzke}
\affil{Max-Planck-Institut f{\"u}r Sonnensystemforschung, Justus-von-Liebig-Weg 3, 37077, G{\"o}ttingen, Germany}

\author{Alexander I. Shapiro}
\affil{Max-Planck-Institut f{\"u}r Sonnensystemforschung, Justus-von-Liebig-Weg 3, 37077, G{\"o}ttingen, Germany}

\author{Savita Mathur}
\affil{Space Science Institute, 4750 Walnut Street, Suite 205, Boulder CO 80301, USA}
\affil{Instituto de Astrof{\' i}sica de Canarias, E-38200 La Laguna, Tenerife, Spain}
\affil{Departamento de Astrof{\' i}sica, Universidad de La Laguna, E-38205 La Laguna, Tenerife, Spain}

\author{Guy R. Davies}
\affil{School of Physics and Astronomy, University of Birmingham, Birmingham B15 2TT, UK}
\affil{Stellar Astrophysics Centre, Department of Physics and Astronomy, Aarhus University, Ny Munkegade 120, 8000, Aarhus C, Denmark}

\author{Mikkel N. Lund}
\affil{School of Physics and Astronomy, University of Birmingham, Birmingham B15 2TT, UK}
\affil{Stellar Astrophysics Centre, Department of Physics and Astronomy, Aarhus University, Ny Munkegade 120, 8000, Aarhus C, Denmark}

\author{Rafael A. Garcia}
\affil{Laboratoire AIM, CEA/DRF -- CNRS -- University Paris 7 Diderot -- IRFU/SAp, Centre de Saclay, 91191, Gif-sur-Yvette, France}

\author{Allan S. Brun}
\affil{Laboratoire AIM, CEA/DRF -- CNRS -- University Paris 7 Diderot -- CEA-Saclay, 91191, Gif-sur-Yvette, France}

\author{David Salabert}
\affil{Laboratoire AIM, CEA/DRF -- CNRS -- University Paris 7 Diderot -- IRFU/SAp, Centre de Saclay, 91191, Gif-sur-Yvette, France}

\author{Pedro P. Avelino}
\affil{Instituto de Astrof{\' i}sica e Ci{\^e}ncias do Espa{\c c}o, Universidade do Porto, CAUP, Rua das Estrelas, PT4150-762 Porto, Portugal}
\affil{Departamento de F{\' i}sica e Astronomia, Faculdade de Ci\^{e}ncias, Universidade do Porto, Rua do Campo Alegre 687, PT4169-007 Porto, Portugal}

\author{Jennifer van Saders}
\affil{Carnegie Observatories, 813 Santa Barbara Street, Pasadena CA 91101, USA}

\author{Ricky Egeland}
\affil{High Altitude Observatory, National Center for Atmospheric Research, P.O. Box 3000, Boulder CO 80307, USA}

\author{Margarida S. Cunha}
\affil{Instituto de Astrof{\' i}sica e Ci{\^e}ncias do Espa{\c c}o, Universidade do Porto, CAUP, Rua das Estrelas, PT4150-762 Porto, Portugal}
\affil{Departamento de F{\' i}sica e Astronomia, Faculdade de Ci\^{e}ncias, Universidade do Porto, Rua do Campo Alegre 687, PT4169-007 Porto, Portugal}

\author{Tiago L. Campante}
\affil{Instituto de Astrof{\' i}sica e Ci{\^e}ncias do Espa{\c c}o, Universidade do Porto, CAUP, Rua das Estrelas, PT4150-762 Porto, Portugal}
\affil{Departamento de F{\' i}sica e Astronomia, Faculdade de Ci\^{e}ncias, Universidade do Porto, Rua do Campo Alegre 687, PT4169-007 Porto, Portugal}

\author{William J. Chaplin}
\affil{School of Physics and Astronomy, University of Birmingham, Birmingham B15 2TT, UK}
\affil{Stellar Astrophysics Centre, Department of Physics and Astronomy, Aarhus University, Ny Munkegade 120, 8000, Aarhus C, Denmark}

\author{Natalie Krivova}
\affil{Max-Planck-Institut f{\"u}r Sonnensystemforschung, Justus-von-Liebig-Weg 3, 37077, G{\"o}ttingen, Germany}

\author{Sami K. Solanki}
\affil{Max-Planck-Institut f{\"u}r Sonnensystemforschung, Justus-von-Liebig-Weg 3, 37077, G{\"o}ttingen, Germany}

\author{Maximilian Stritzinger}
\affil{Department of Physics and Astronomy, Aarhus University, Ny Munkegade 120, 8000, Aarhus C, Denmark}

\author{Mads F. Knudsen}
\affil{Department of Geoscience, Aarhus University, H{\o}egh-Guldbergs Gade 2, 8000, Aarhus C, Denmark}

\begin{abstract}
Observations of Sun-like stars over the last half-century have improved our understanding of how magnetic dynamos, like that responsible for the 11-year solar cycle, change with rotation, mass and age. Here we show for the first time how metallicity can affect a stellar dynamo. Using the most complete set of observations of a stellar cycle ever obtained for a Sun-like star, we show how the solar analog HD~173701 exhibits solar-like differential rotation and a 7.4-year activity cycle. While the duration of the cycle is comparable to that generated by the solar dynamo, the amplitude of the brightness variability is substantially stronger. The only significant difference between HD~173701 and the Sun is its metallicity, which is twice the solar value. Therefore, this provides a unique opportunity to study the effect of the higher metallicity on the dynamo acting in this star and to obtain a comprehensive understanding of the physical mechanisms responsible for the observed photometric variability. The observations can be explained by the higher metallicity of the star, which is predicted to foster a deeper outer convection zone and a higher facular contrast, resulting in stronger variability.
\end{abstract}

\section{Introduction}
The number of spots on the surface of the Sun changes over a characteristic 11-year cycle, and this sunspot cycle is accompanied by a $\sim$0.1\% change in brightness \citep{2009A&A...501L..27F}. The increase in brightness with increasing spot coverage over the solar cycle is due to the compensating effect of faculae \citep{2006Natur.443..161F}. The number of sunspots is also known to vary on much longer timescales, with episodes of complete disappearance like the 17th century Maunder Minimum \citep{1976Sci...192.1189E}. It remains a matter of debate how bright the Sun was during the Maunder Minimum \citep[see][for a recent review]{2013ARA&A..51..311S}. In fact, we do not know how the Sun's brightness changes on timescales longer than a few decades. One way to improve this situation is to measure analogous brightness variations in Sun-like stars and use such measurements to reveal the relationship between spots, magnetic activity and brightness changes on different timescales. This was first done by \citet{1992Natur.360..653L}, who used 8 years of observations at the Lowell observatory of 33 Sun-like stars to conclude that ``...the Sun is in an unusually steady phase compared to similar stars, which means that reconstructing the past historical brightness record, for example from sunspot records, may be more risky than has been generally thought." This conclusion was challenged by \citet{2009AJ....138..312H}, who argued that the Sun's apparently low brightness variability compared to other Sun-like stars was due to selection effects. The only other inactive star in the ensemble, the solar twin 18 Sco, has a brightness variability lower than the Sun \citep{2007AJ....133.2206H}.

In order to use Sun-like stars to reconstruct the historical brightness variability of the Sun, the fundamental properties of the stars should be carefully analysed. It is particularly important to understand whether the dynamo in the stars has the same nature as the solar dynamo. The best tool for such an analysis is asteroseismology, where the eigenfrequencies of the stars may be used to accurately determine fundamental stellar properties like radius, mass, age, composition and rotation period \citep[see e.g.][]{2012ApJ...748L..10M, 2015MNRAS.446.2959D,2017ApJ...835..172L}.

Here we analyse the Sun-like star HD~173701 (KIC 8006161), which is one of the brightest stars observed by {\it Kepler} and therefore also among the stars with the best known fundamental properties like radius, mass and age as they have been measured with asteroseismology (see Table~1). The asteroseismic analysis reveals that HD~173701 is almost identical to the Sun with respect to radius, mass and age, but it has a metallicity that is twice as high as the solar value. Assuming everything else equal, HD~173701 therefore allows us to measure the effect of metallicity on stellar cycles. The nature of our study is therefore fundamentally different from the ensemble studies by e.g. \citet{1992Natur.360..653L, 1998ApJS..118..239R, 1995ApJ...438..269B, 1996AJ....111..439H, 2005PASP..117..657W}, where a large number of stars were analysed whose fundamental properties are not well constrained. Here we only study one star, but its fundamental properties are extremely well determined.

\begin{table}[!b]
\caption{Stellar parameters for HD~173701. $^{*}$ from \citet{2016arXiv161208990C} and $^{**}$ from \citet{2015ApJ...808..187B}.}
\centering
\begin{tabular}{lrl}
\hline \hline
Radius$^*$: & $0.930 \pm 0.009$ R$_{\odot}$\\
Mass$^*$: & $1.00 \pm 0.03$ M$_{\odot}$\\
Log g$^*$: & $4.498 \pm 0.003$ \\
Age$^*$: & $4.57 \pm 0.36$ Gyr$$\\
Effective temperature$^{**}$ & 5488 $\pm$ 77 K\\
Metallicity$^{**}$: & 0.3 $\pm$ 0.1\\
Rotation period: &$ 21_{-2}^{+2}$ days\\
Inclination: & $38_{-4}^{+3}$ degrees\\
Cycle period: & $7.41 \pm 1.16$ years\\
\hline
\end{tabular}
\label{tab2}
\end{table}

Asteroseismology not only allows us to measure the fundamental parameters of the stars, it also allows us to investigate the relation between the cycle related phenomena taking place inside the stars to those taking place on the surface. HD~1737101 is a perfect candidate for such a study, not only because we can determine the temporal variability of the eigenfrequencies, but also because we have many different measurements of cycle related phenomena. This allows us to make a detailed comparison between the dynamos operating in HD~173701 and in the Sun, and how the cycle manifests itself on the surfaces of the two stars.

The paper is arranged as follows: in Section 2, we describe the different analyses we conduct on HD~173701 including: spectroscopy, photometry and asteroseismology. In Section 3, we compare the results of our analysis with a similar analysis of the Sun, to investigate the differences between HD~173701 and the Sun. In Section 4, we discuss the implications of these results for our understanding of the variability in HD~173701. 

\section{Analysis}
The analysis of HD~173701 consists of a spectroscopic, a photometric and an asteroseismic analysis, as well as an analysis of the photospheric activity proxy. A comparable analysis of the Sun is performed to determine the impact of the higher metallicity of HD~173701.

\subsection{Spectroscopy} 
In the Mount Wilson HK project, chromospheric emission was measured with the dimensionless S index \citep{{1991ApJS...76..383D}}:

\begin{equation}
S=\alpha \cdot \frac{H+K}{R+V},
\end{equation}
where $H$ and $K$ are the recorded counts in 1.09 {\AA} full-width at half-maximum triangular bandpasses centered on the Ca~{\sc ii} $H$ and $K$ lines at 396.8 and 393.4 nm, respectively. $V$ and $R$ are two 20 {\AA} wide reference bandpasses centered on 390.1 and 400.1 nm, respectively, while $\alpha$ is a normalization constant. 

Measured S indices for almost 2300 stars, including HD~173701, from the Mount Wilson HK project are available for  download from the National Solar Observatory webpage\footnote{\url{ftp://solis.nso.edu/MountWilson_HK/}}. These observations include 3 measurements in 1978, 165 in 1983 and 24 in 1984.  Based on an ensemble of flat-activity stars, \citet{1995ApJ...438..269B} estimated a nightly measurement uncertainty of 1.2\%.

From the Nordic Optical Telescope (NOT) we obtained 12 epochs of observations from 2010 to 2014. These were reduced as described in \citet{2013MNRAS.433.3227K, 2009MNRAS.399..914K} and note that the data set does now include observations from 2013.

The normalization constant ($\alpha$) is usually obtained by measuring a number of stars that were part of the Mount Wilson HK project. The calibration does not have to be linear \citep{2010ApJ...725..875I}. This approach was, however, not possible in the study by \citet{2013MNRAS.433.3227K}, as only one star was available for comparison. Instead, the excess flux, defined as the surface flux arising from magnetic sources, was measured. The excess flux was then used to calculate a pseudo-S index by calibration with the effective temperature \citep{2013MNRAS.433.3227K}. 

The observations from the Keck telescope were presented by \citet{2010ApJ...725..875I} and we include additional observations from 2015. A calibration of the S indices was obtained from 151 stars that are both part of the Mount Wilson HK project and the California Planet Search program \citep{2005PASP..117..657W, 2010ApJ...725..875I}. Uncertainties, were calculated as described in \citet{2010ApJ...725..875I}.

Using 21 stars observed with both the NOT and Keck telescopes, the instrumental S indices from the former were calibrated using the calibrated S indices from the latter telescope. In order to minimise numerical effects in the calculation of the S indices of HD~173701, all spectra from the NOT and Keck telescope were reanalysed using the same code. The uncertainties of the NOT measurements were obtained using nights with multiple observations to obtain the following relation between the uncertainty of the mean value of the chromospheric activity measured that night and S/N: $\sigma = 0.011/\sqrt{\rm S/N}$. An additional flat noise term of 0.002 was added in quadrature to the uncertainty of the mean values \citep{2011arXiv1107.5325L}.

Combining the observations from the NOT and Keck telescopes with annual average observations from the Mount Wilson HK Project a cycle period of $7.41 \pm 1.16$ years is obtained using least squares, thereby overlapping with the period covered by the nominal {\it Kepler} mission \citep{2010ApJ...713L..79K}, as demonstrated in Fig.~1. The chromospheric emission of HD~173701 shows a cyclic variability that is 2.2--2.7 times stronger than that of the Sun (Fig.~2).
\begin{figure}[!t]
\plotone{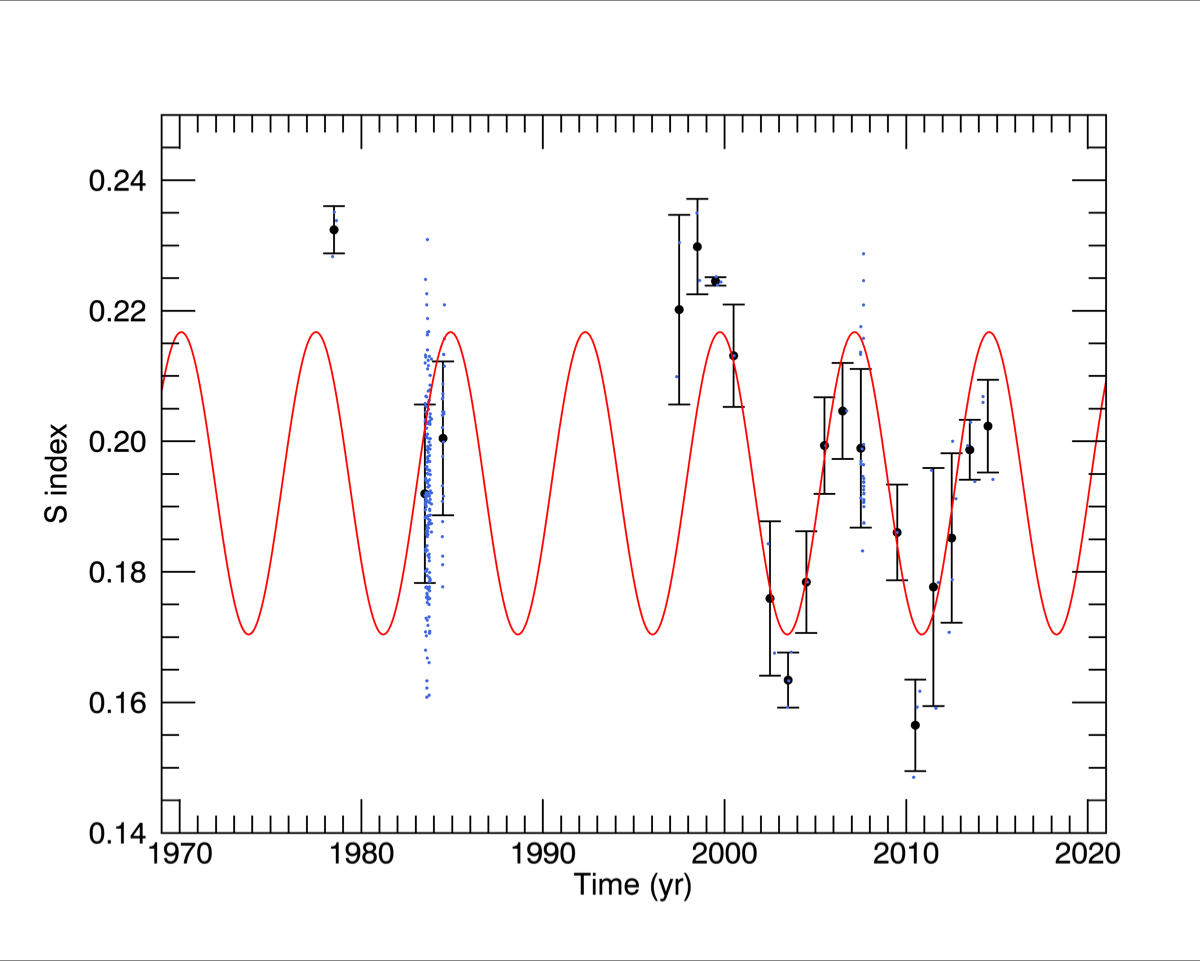}
\caption{Modeling the cycle in HD~173701. The red curve is a least squares fit to annual means (black points with error-bars) of all the available observations (blue crosses) with a period of $7.41\pm1.16$  years.}
\end{figure}

\begin{figure*}[!t]
\plotone{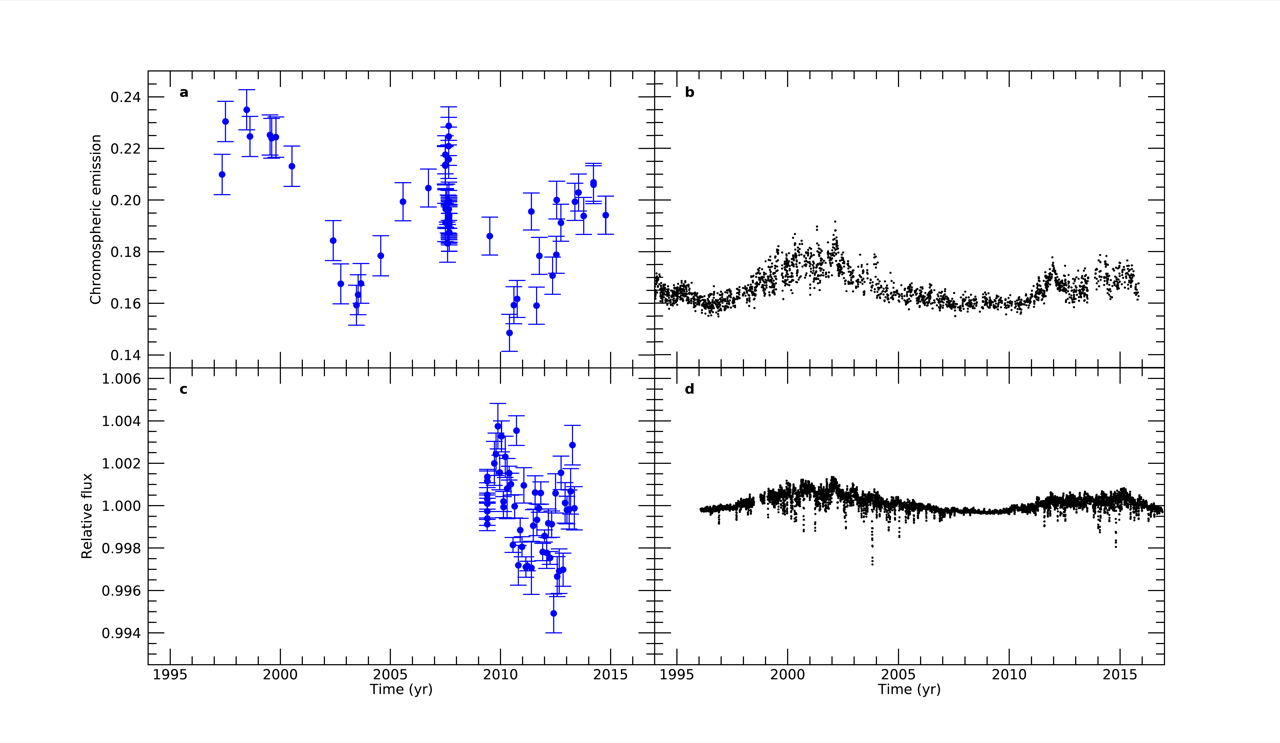}
\caption{The stellar cycle in HD~173701 compared to the Sun. Panels show the cycles in HD~173701 (left) and the Sun (right) as seen in the chromospheric emission ({\bf a} and {\bf b}) as well as the response to this magnetic cycle in the relative photometric flux of the stars ({\bf c} and {\bf d}). The scale of the axes is the same for HD~173701 and the Sun. A clear 7.4-year cycle is seen in the chromospheric emission of HD~173701 ({\bf a}). The cycle is superimposed on observations extending back to 1978 in Fig. 1. It is seen that the relative photometric flux follows the rising phase of the last cycle seen in the chromospheric emission.}
\end{figure*}

\subsection{Photometry} 
During the primary {\it Kepler} mission, the telescope recorded aperture photometry of almost 200,000 stars at 30-minute cadence \citep{2010Sci...327..977B, 2010ApJ...713L..87J}. As the recorded apertures are smaller than the typical point spread function (PSF) for the telescope, small changes in the telescope position, temperature, or PSF overwhelm small changes in the intrinsic brightness of the star, making long-term brightness variations inaccessible with standard {\it Kepler} photometry.
These variations can be recovered through the Full Frame Images (FFIs), in which the entire {\it Kepler} detector was recorded and sent to Earth approximately monthly throughout the mission, providing an opportunity to measure aperture photometry for each isolated star over its entire PSF.

We use the f3 software package described in \citet{2017arXiv170507928M} to infer the brightness of HD 173701 and 15 bright, nearby comparison stars in 52 FFIs spanning the Kepler mission. All target stars fall within 15$^{''}$ of HD 173701. We inspect each light curve by eye to ensure none of the comparison stars are intrinsically variable.
For each of the four orientations of the {\it Kepler} telescope, we then measure the flux for HD~173701 relative to each of the comparison stars, building a time series in observed brightness that accounts for instrumental systematics.
The orientations are considered separately as the underlying flat field is poorly understood, so the percent-level inter- and intra- pixel sensitivity changes across the detector can induce an artificial offset from orientation to orientation. Finally, the four sets of data are combined into one by dividing by the median flux value in each orientation, and applying a linear offset to all data in each individual orientation such that the residuals of a quadratic fit to the data are minimized. 

The photometric uncertainties are calculated from the quadratic fit as described in \citet{2017arXiv170507928M}. This approach assumes that HD 173701, which is heavily saturated in the FFIs, behaves similar to the non-saturated reference stars. This assumption is supported by the fact that we are able to obtain nearly Poisson limited photometry for stars brighter than HD 173701 \citep{2010ApJ...713L.160G}.

Our analysis shows that the broad-band photometric variability follows the cyclic variability seen in the chromospheric emission (Fig.~2). The standard deviation of the photometric variability of HD~173701 is 2.4--4.8 times larger than the photometric variability observed in the Sun (see Section 3 for a detailed explanation of how both the lower and the upper boundaries were obtained). The relative level of solar photometric variability compared to other Sun-like stars is an important parameter in many Sun-climate studies. The high value of the photometric variability we find for this Sun-like star indicates that the Sun shows unusually weak photometric variability, supporting the conclusion of \citet{1992Natur.360..653L} \citep[see also][]{2007ApJS..171..260L}. However, given the sample size of one, caution should be taken when drawing any such conclusions from our results.

\subsection{Asteroseismology}\label{sec:seismic} 
The asteroseismic analysis in this study has three purposes. Firstly, we use the results from the asteroseismic analysis by \citet{2016arXiv161208990C} and adopt the general parameters in Table~1 (note that the last two are spectroscopic parameters from \citet{2015ApJ...808..187B}, as asteroseismology is generally not very efficient in constraining effective temperature and metallicity). We note that the parameters in Table~1 result in a luminosity that is around $2\sigma$ higher than what is found by Hipparcos and Gaia, the luminosity were however not used in the asteroseismic analysis by \citet{2016arXiv161208990C}. Tests have showed that including the luminosity in the asteroseismic analysis leads to an insignificant higher mass ($\sim1.02$ M$_{\odot}$). Secondly, we measure the rotation rate and inclination of the star and thirdly, we use asteroseismology to we measure the effect of the activity cycle on the temporal evolution of the eigenfrequencies of the star.

The raw data for the asteroseismic analysis were taken from the {\it Kepler} Asteroseismic Science Operations Center (KASOC) and corrected using the KASOC filter \citep{2014MNRAS.445.2698H}. 

Information on stellar rotation can be extracted from the measurable properties of rotationally split non-radial eigenfrequencies in the frequency power spectrum of main-sequence stars \citep{2013ApJ...766..101C, 2014MNRAS.444.3592D, 2015MNRAS.446.2959D,2016ApJ...819...85C}. Estimates of rotational properties are the outputs of so-called {\it peak-bagging} procedures and here we use the methods of \citet{ 2015MNRAS.446.2959D, 2016MNRAS.456.2183D} to determine $\left( \nu_{\rm s} \sin i, i, \nu_{\rm s}, P \right)$, where $i$ is the angle of inclination, $\nu_{\rm s}$ is the rotational splitting in frequency, and $P$ is the average asteroseismic rotation period.

This analysis returns an equatorial rotation period of $21_{-2}^{+2}$ days and an inclination of $38_{-4}^{+3}$ degrees (Fig.~6). The reliability of the asteroseismic method for measuring rotation and inclination was tested using the algorithm developed by \citet{2017ApJ...835..172L}. This test gave very consistent results with a rotation period of $21_{-5}^{+4}$ days and an inclination of $37_{-8}^{+6}$ degrees. The rotation period we find with asteroseismology is slightly shorter than the period we find for the activity modulated signal in the photometry (25-35 days, see Section 3.2). The reason for this is likely that the asteroseismic signal mainly originates from the equatorial regions whereas the activity modulated signal could originate from higher latitudes. The Sun-as-a-star seismic synodic rotation period of the Sun is 26.9 days \citep{2014MNRAS.439.2025D}, comparable to the solar equatorial rotation period. A slightly larger value, but still within the uncertainties, was found by \citet{2015MNRAS.446.2959D}.

\begin{figure}[!t]
\plotone{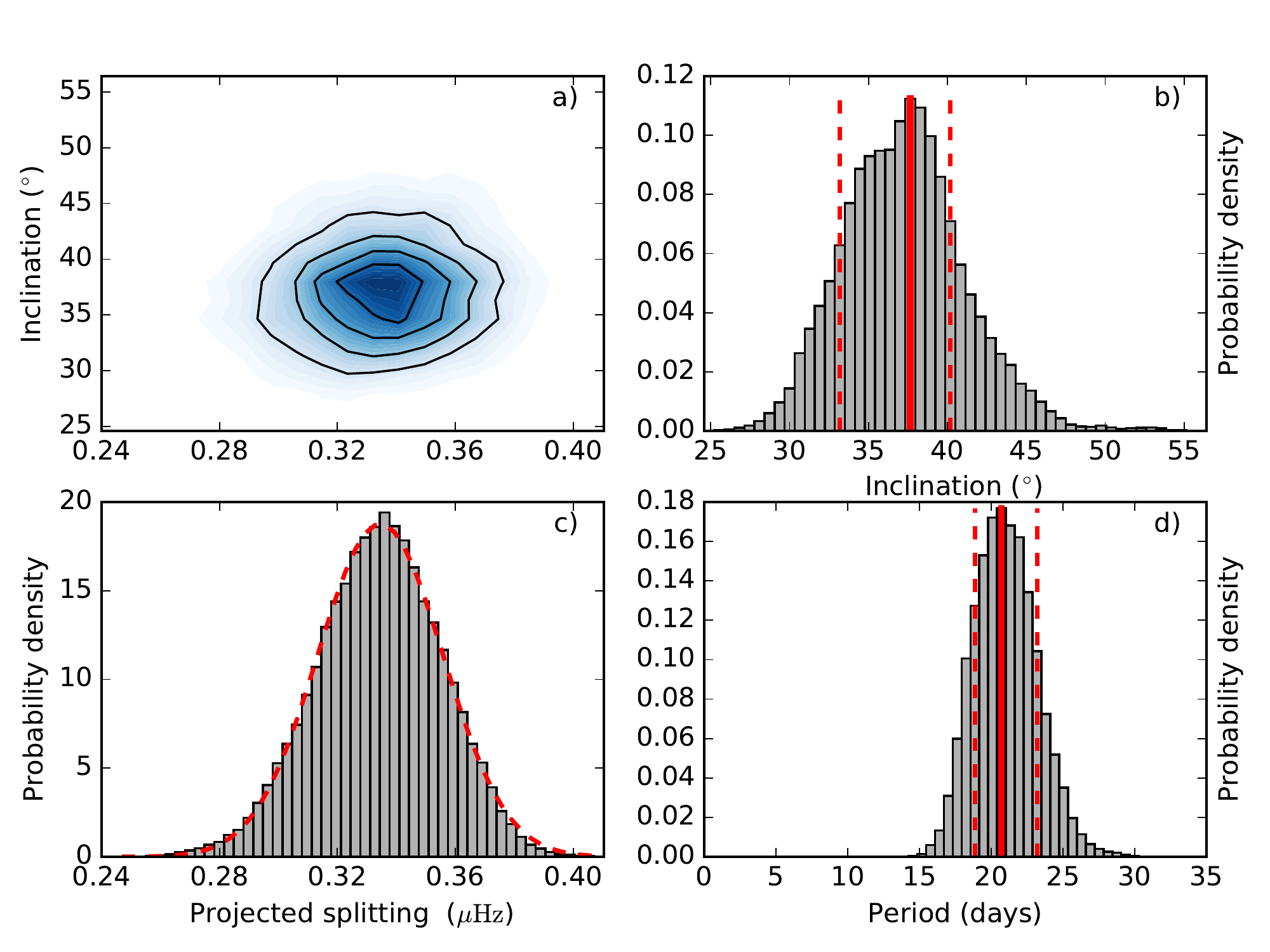}
\caption{Marginalised posterior probability distributions for the rotational induced signal in the eigenfrequencies of HD~173701. The plot is composed of inclination versus projected splitting ({\bf a}), inclination ({\bf b}), projected splitting ({\bf c}) and period of rotation ({\bf d}). For a discussion of the method see \citet{2015MNRAS.446.2959D}.}
\end{figure}

For the analysis of the temporal evolution of the eigenfrequencies, the processed time series was segmented into 90-day-long sub-series with an overlap of 45 days. The corresponding frequency power spectra were then obtained from the periodogram of each sub-series.

To describe the background signal, we use three components: (i) an exponential decay of active regions \citep[e.g.,][]{Garcia2009,2016ApJ...819...85C}; (ii) a Harvey-like profile for the granulation \citep[e.g.,][]{Harvey1985}; (iii) and a constant offset denoting the photon noise. The background parameters --- corresponding to the best fit to the power spectra --- are then fixed for the peak-bagging analysis.
\begin{figure}[!t]
\plotone{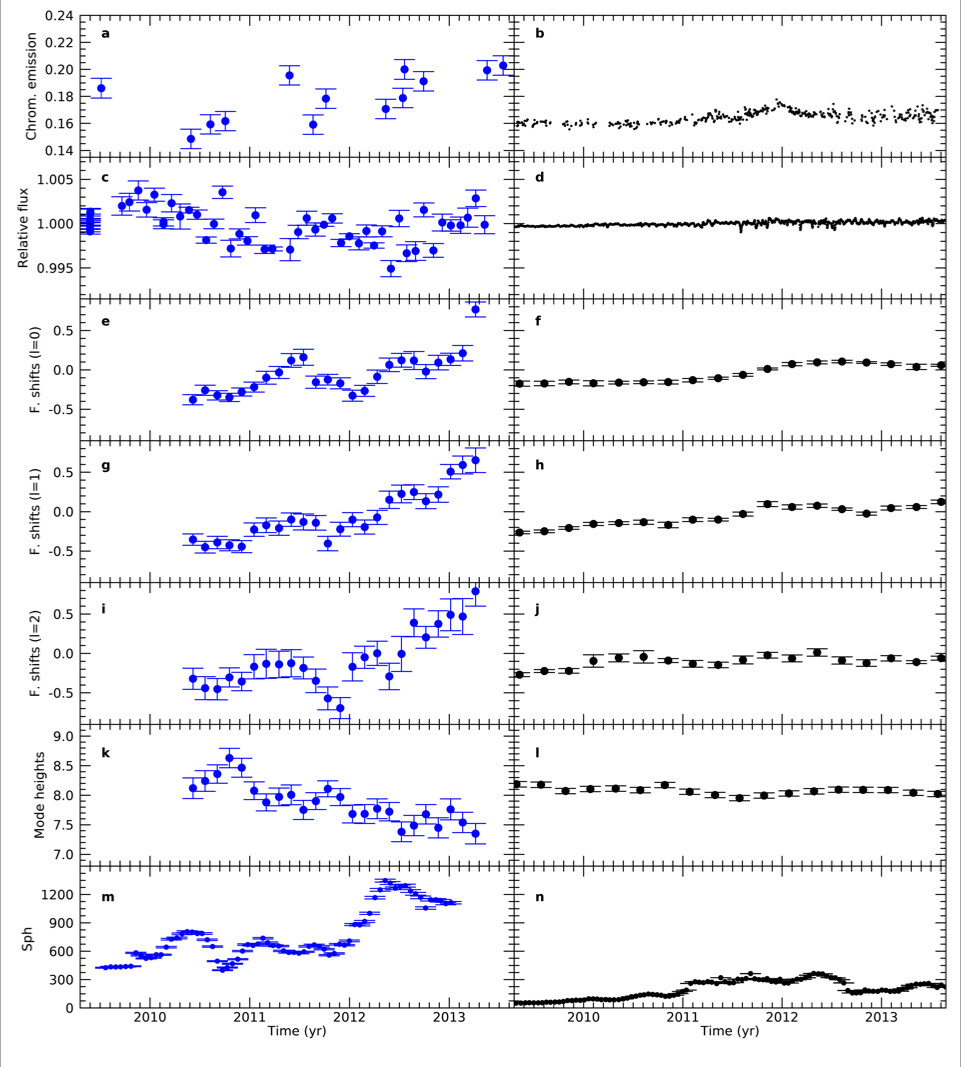}
\caption{The rising phase of the last cycle in HD~173701 compared to the Sun.  The panels show the chromospheric emission ({\bf a} and {\bf b}), the relative flux ({\bf c} and {\bf d}), radial frequency shifts ({\bf e} and {\bf f}), dipolar frequency shifts ({\bf g} and {\bf h}), quadrupolar frequency shifts ({\bf i} and {\bf j}), logarithmic mode heights of the eigenfrequencies ({\bf k} and {\bf l}) and photospheric activity proxy ({\bf m} and {\bf n}).}
\end{figure}

To perform a global fit to the oscillation modes and estimate the respective model parameters (eigenfrequencies, as well as heights and linewidths of the radial modes, rotational splitting, and stellar inclination angle), we followed a Bayesian approach \citep[e.g.,][]{2011A&A...534A...6C,2011A&A...527A..56H,2016MNRAS.456.2183D,2017ApJ...835..172L} through implementation of an affine-invariant MCMC ensemble sampler \citep{Goodman2010,2013PASP..125..306F}. The Bayesian peak-bagging analysis is fully described in Santos et al.~(in prep.). In summary, we apply uniform priors to the mode frequencies \citep[within $8\,{\rm\mu Hz}$ of the eigenfrequencies determined by][]{2017ApJ...835..172L}, linewidths and heights. Following the approach of \citet{2016MNRAS.456.2183D}, we also apply priors to the large and small frequency separations. Finally, we use the posterior probability distributions obtained by \citet{2017ApJ...835..172L} to define the prior probabilities on the rotational splitting and inclination angle.

Once the eigenfrequencies for all sub-series have been estimated, we compute the weighted mean frequency shifts and corresponding uncertainties as:
\begin{equation}
\delta\nu_l(t)=\frac{\sum_{n}\delta\nu_{nl}(t)/\sigma^2_{nl}(t)}{\sum_{n}1/\sigma^2_{nl}(t)},
\end{equation}
and
\begin{equation}
\sigma_l(t)=\left[\sum_{n}1/\sigma_{nl}^2(t)\right]^{-1/2},
\end{equation}
where $\delta\nu_{nl}(t)$ corresponds to the variation in frequency of a mode of radial order $n$ and angular degree $l$ with respect to the average value, and $\sigma_{nl}(t)$ is the respective uncertainty. 

The temporal variability of radial ($l$=0), dipolar ($l$=1) and quadrupolar ($l$=2) oscillation modes averaged over the five central radial orders closely follow the cyclic variability seen in the chromospheric emission, and show the same characteristic behaviour as seen in the Sun (Fig.~4~\&~5), i.e., the standard deviation increases with higher degree $l$ (though the increase is only marginal between the $l$=1 and $l$=2 modes). This suggests that the origin of the perturbation to the frequencies is located in the outer layers of the star and thus the dynamo driving the variability in HD~173701 is similar to the dynamo driving the solar cycle.

The 11-year solar cycle can also be seen in the height of the oscillation modes in a power spectrum \citep{2000MNRAS.313...32C}. Since mode heights are approximately distributed according to a log-normal distribution, we use the logarithm of the mode heights and proceed in the same manner as for the frequency shifts in computing the mean logarithmic mode heights. The standard deviation of the mode height variations in HD~173701 is 1.6--2.4 times larger than those observed for the Sun. 

\begin{figure}[!t]
\plotone{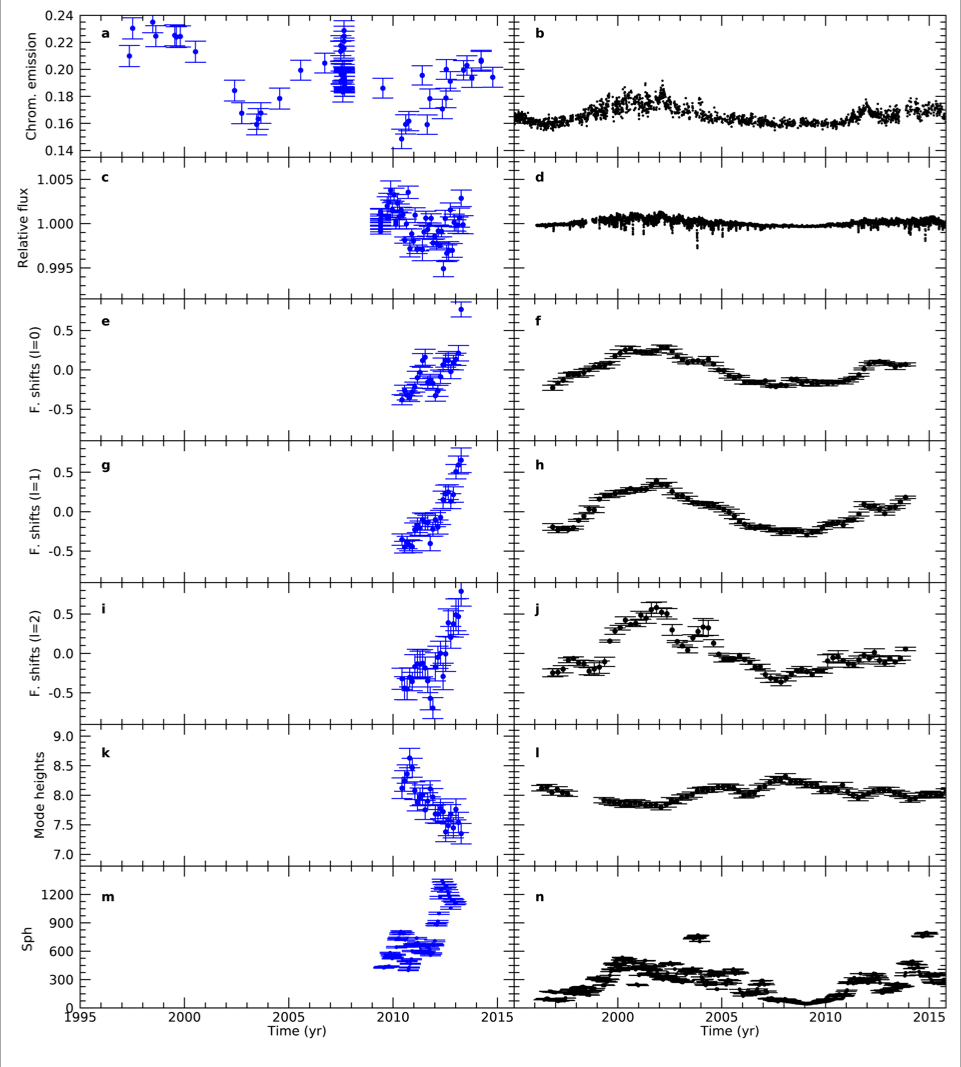}
\caption{Two full cycles. The figure shows the same as Fig.~4, but here zoomed out to see the last solar cycle.}
\end{figure}

\begin{table*}
\begin{footnotesize}
\caption{Standard deviations}
\begin{tabular}{lrrrrr}
\hline \hline
&
HD~173701 &
\multicolumn{2}{c}{The Sun} &
\multicolumn{2}{c}{Ratio} \\
\hline
&Original & Original & Adjusted & Original & Adjusted \\
\hline
Chromospheric emission: & $0.0194\pm0.0011$ 	& 0.0072 	& $0.0090\pm0.0008$ 	& 2.7		& 2.2\\
Relative flux: 			& $0.0019\pm0.0001$ 	& 0.0004	& $0.0008\pm0.0001$	& 4.8		& 2.4\\
Frequency shifts (l=0) [$\mu$Hz]: 	& $0.2547\pm0.0169$	& 0.16 	& $0.1781\pm0.0175$	& 1.6		& 1.4\\
Frequency shifts (l=1) [$\mu$Hz]: 	& $0.3273\pm0.0217$ 	& 0.22 	& $0.2349\pm0.0249$	& 1.5		& 1.4\\
Frequency shifts (l=2) [$\mu$Hz]: 	& $0.3685\pm0.0369$	& 0.28 	& $0.3186\pm0.0380$	& 1.3		& 1.2\\
Mode heights [log ppm$^2/\mu$Hz] :			& $0.3374\pm0.0314$	& 0.14	& $0.2121\pm0.0304$	& 2.4		& 1.6\\
Photospheric proxy [ppm]: 		& $273.89\pm1.37$ 		& 160.22	& $160.19\pm12.65$		& 1.7		& 1.7\\
\hline
\end{tabular}
\label{tab2}
\end{footnotesize}
\end{table*}

\subsection{The photospheric activity proxy}\label{sec:sph}
The photospheric activity proxy, $S_\mathrm{ph}$, is a measurement of stellar magnetic variability derived by means of the surface rotation, $P_\mathrm{rot}$ \citep{2014A&A...572A..34G, 2015A&A...583A.134F, 2016A&A...596A..31S}. 
The $S_\mathrm{ph}$ proxy is defined as the mean value of the light curve fluctuations estimated as the standard deviations calculated over sub-series of length $5 \times P_\mathrm{rot}$. 
In this way, \citet{2014A&A...562A.124M} demonstrated that most of the measured variability is only related to the magnetism (i.e. the spots and faculae) and not to the other sources of variability at different timescales, such as convective motions, oscillations, stellar companion, or instrumental problems. This assumes that the spots and faculae are not distributed close to the equator, but at higher latitudes. Otherwise, the value of $S_\mathrm{ph}$ obtained would have been a lower limit of the true photospheric variability and the rotation signature would have been difficult to measure. This implies that the rotation period found in the activity modulation from photometry reflects mid to low latitudes, which would then be slower on HD~173701 than on the Sun. The error on $S_\mathrm{ph}$ was returned as the standard error of the mean value. The $S_\mathrm{ph}$ was measured on Kepler light curves calibrated with the KADACS software as described in \citet{2011MNRAS.414L...6G} and \citet{2015A&A...574A..18P}

The measured $S_\mathrm{ph}$ values are shown in Figs.~4~\&~5. The measured $S_\mathrm{ph}$ values of HD~173701 show a standard deviation 1.7 times higher than what is observed for the Sun.

\subsection{Solar data}
We used the compilation of the total solar irradiance (TSI) from \citet{2009A&A...501L..27F} to compare the photometric variability of HD~173701 to the Sun over the considered period, these data represent a composite from the DIARAD and PMO6V absolute radiometers on the {\it Solar and Heliospheric Observatory} (SOHO) and have a bolometric character comparable to the very broad visible bandpass (423--897 nm) used by {\it Kepler} \citep{2013ApJ...769...37B}.

For the solar S indices, observations of the daily K-index $K_{\rm SP}$ from the Evans Coronal Facility at Sacramento Peak \citep{1998ASPC..140..301K} were transformed to the MWO S-index scale using the calibration of \citet{2017ApJ...835...25E}. This calibration was developed using overlapping observations (1993-2003) of reflected sunlight from the Moon through the MWO HKP-2 spectrophotometer.

In order to compare the asteroseismic signal in HD~173701 to the Sun, we used the time-dependent measurements of solar eigenfrequencies by \citet{2015A&A...578A.137S}. The values in Figs.~4~\&~5 were calculated as mean values of the five central orders ($n=21-25$) and the errorbars represent the uncertainties on the mean values. The logarithmic mode heights of the eigenfrequencies were also measured as described in \citet{2015A&A...578A.137S}, but here observations from the {\it Variability of solar IRradiance and Gravity Oscillations} (VIRGO) instrument on SOHO were used.

\section{Results}
\subsection{Amplitude of the cyclic variability}
The amplitude of the activity cycle in HD~173701 is compared to the solar cycle in Figs.~4~\&~5. We first calculate the standard deviation of the different cycle related parameters we have measured in HD~173701 (chromospheric emission, relative photometry, frequency shifts, mode heights and the photospheric activity proxy). The uncertainties on the calculated standard deviations are calculated by a bootstrap test, where the measurements are randomly shifted according to the measured uncertainty of the individual measurements. We then calculated the standard deviation of the solar parameters over a solar cycle. These values we refer to as original values (Table~2). The calculated standard deviations are affected by both the sampling and the uncertainties on the measurements. We therefore also calculate the standard deviations of the solar parameters by randomly selecting as many solar measurements as we have stellar measurements over a solar cycle and adding a normally distributed noise term given by the uncertainty of the stellar observations. These values are referred to as adjusted values (Table 2). The adjusted standard deviations for the solar parameters are generally higher than the original standard deviations and they are highly sensitive to the absolute level of the uncertainties on the stellar measurements. This is especially clear for the standard deviation of the relative photometry, where the adjusted standard deviation is twice as large as the original. 

The uncertainties on the relative photometry on the FFIs are due to a number of noise sources, where the most important are: photon noise, variability of comparison stars and inter- and intra-pixel sensitivity changes \citep{2017arXiv170507928M}. For HD~173701 the largest contribution is likely to come from percent-level inter- and intra-pixel sensitivity changes across the detector that can induce artificial offsets when the spacecraft changes orientation every quarter. These changes will introduce slow drifts in the measured photometry, that are likely not to have a large effect on the standard deviation. This suggests that adjusted standard deviation of the solar photometry should be seen as a upper limit on the standard deviation and that the standard deviation to compare with the stellar result is closer to the original standard deviations.

In addition to this, the relative solar photometry is given as daily averages \citep{2009A&A...501L..27F}, whereas the stellar measurements are obtained over only half an hour. On time-scales less than a day, the solar photometry is dominated by granulation and oscillations, which is not the focus of our studies. We used the atmosphere model described below to calculate the difference in the standard deviation of the solar photometry calculated from either 30-minutes or 24-hour averages. Here the former turned out to be only 0.1\% larger. We therefore adopt this value to be insignificant.

\subsection{Peak-height ratios}
Starspot modulation of the stellar light curves encodes information about the stellar rotation and magnetic activity (see also Section \ref{sec:sph}). If the star is differentially rotating, spots at different latitudes may have different rotation rates.

Based on the periodogram analysis, and in particular on the ratios between the heights of the second and first harmonics of the rotation period (hereafter peak-height ratios), \citet{Reinhold2015} proposed a method to determine the sign of differential rotation. With a detailed analysis of synthetic data, \citet{Santos2017} showed that the method was not fully valid, in some cases, leading to false-positives/negatives of the sign of differential rotation \citep[for details see][]{Santos2017}. Nevertheless, \citet{Santos2017} also showed that the peak-height ratios may provide a simple and fast way to constrain stellar inclination and spot latitude. Namely, if the stellar inclination is known (e.g. from asteroseismology; Section \ref{sec:seismic}), one may estimate the spot latitudes.

\begin{figure}[!t]
\plotone{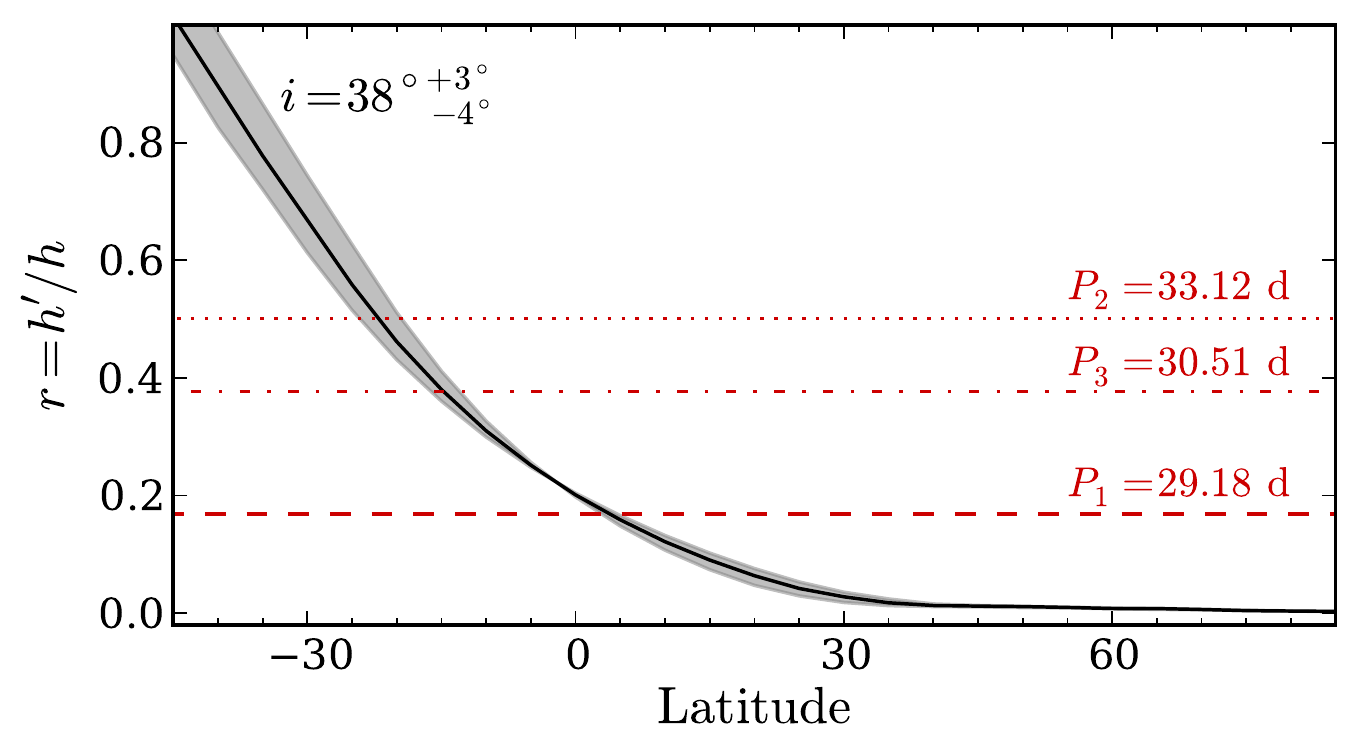}
\caption{Comparison between the theoretical peak-height ratios for $i=38^{\circ+3^\circ}_{\,\,\,-4^\circ}$ (solid black line and gray region) and the values recovered from the periodogram analysis of the {\it Kepler} light curve of HD~173701 (dashed lines).}\label{fig:phratios}
\end{figure}

Fig. \ref{fig:phratios} compares the theoretical latitude-ratio relation \citep[following the approach in][]{Santos2017} with the peak-height ratios obtained from the periodogram analysis of the full {\it Kepler} light curve for HD~173701. We repeat the same analysis for two independent sub-series of 735 days.

\begin{figure}[!t]
\plotone{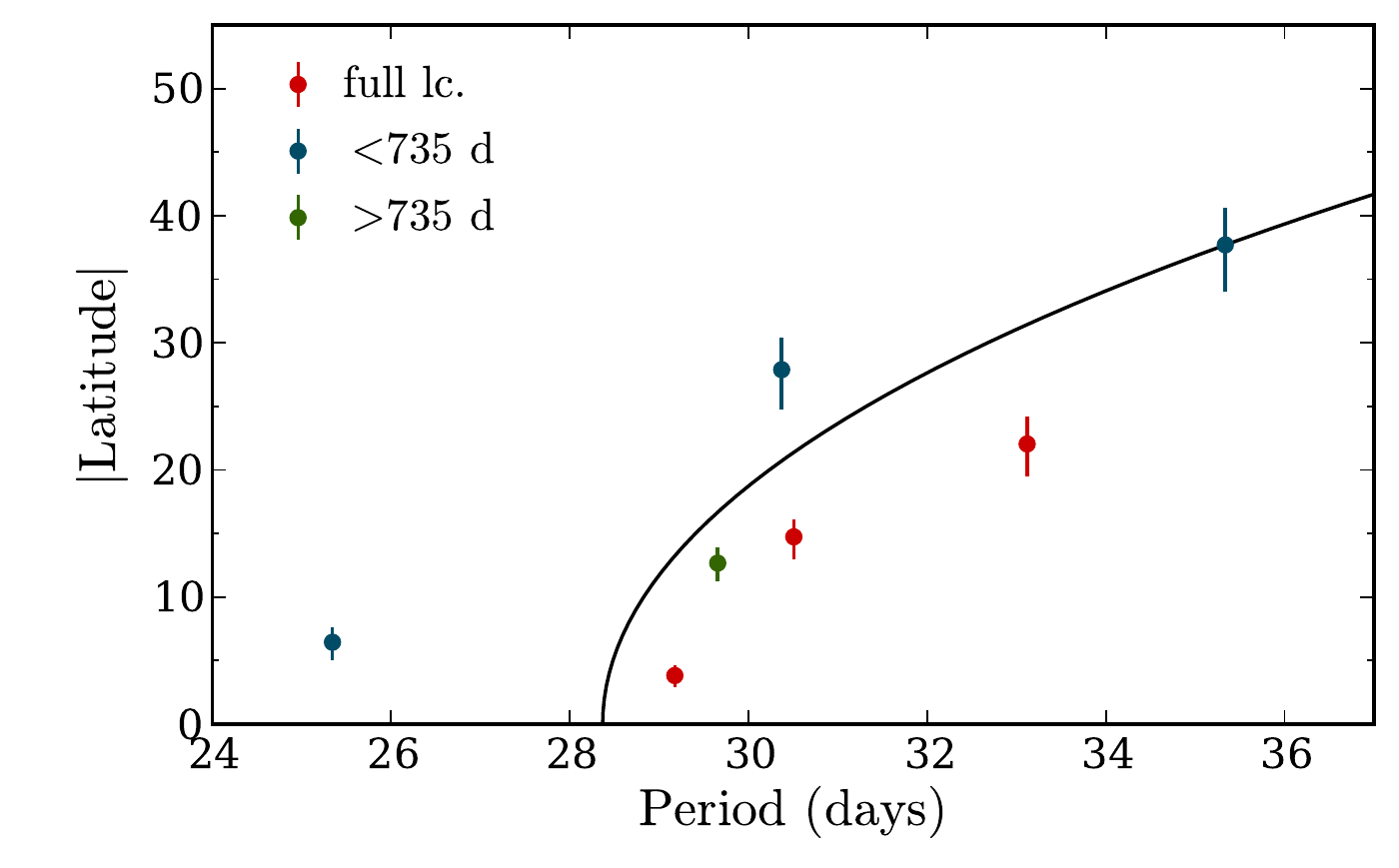}
\caption{Absolute spot latitudes, inferred from the peak-height ratios, as a function of the rotation period. The dashed, dotted, and dash-dotted lines correspond to the results from the full light curve, first 735~d, and last 735~d, respectively. The solid line shows the best fit obtained with the rotation profile in equation~\ref{eq:rotprof}, where $P_{\rm eq}=28.37$ d and $\alpha=0.53$.}\label{fig:rotprof}
\end{figure}

Having the peak-height ratios, we infer the spot latitudes from the theoretical relation. Fig. \ref{fig:rotprof} shows the absolute values of the inferred spot latitudes as a function of the rotation period. The error bars on the spot latitudes are based on the uncertainty on the stellar inclination. The solid line shows the best fit obtained with a rotation profile of the form 
\begin{equation}
P(L)=\frac{P_{\rm eq}}{1-\alpha\sin^2L},\label{eq:rotprof}
\end{equation}
where $P(L)$ is the rotation period at a given latitude $L$, $P_{\rm eq}$ corresponds to the rotation period at the equator, and $\alpha$ is related with the surface shear. The parameters of the best fit are $P_{\rm eq}=28.37\,\rm d$ and $\alpha=0.53$. The rotational profile for HD~173701 suggests strong solar-like differential rotation (equator rotates faster than the poles). For comparison, the solar relative shear is about $\alpha_\odot\sim0.15$ \citep[e.g.][]{Snodgrass1983,Snodgrass1990,1996ApJ...466..384D}. The periodogram analysis and, in particular, the peak-height ratios may be affected by spot evolution, which is not considered here. However, given that the star is significantly active, one may expect spots to be long-lived and their evolution to play a modest role in the present case. The fact that the results we obtain from the peak-height ratios are consistent with the results from the complementary analyses presented in this work also seems to indicate that that is the case.

\subsection{Differential rotation}
Using the $Kepler$ photometry we obtain a rotation period varying between 25 and 35 days (also consistent with the results in Fig. \ref{fig:rotprof}) and a mean value of 27 days by repeatedly computing the Lomb-Scargle periodogram on a 200-day sliding window, following \citet{1995SoPh..159...53D}. This value is in reasonable agreement with the value of $29.8 \pm 3.1$ days obtained by \citet{2014A&A...572A..34G}. The asteroseismic analysis returns a rotation period of $21_{-2}^{+2}$ days. Keeping in mind that asteroseismic measurement mainly reflects the equatorial rotation period in the outermost layers of the star \citep{2014ApJ...790..121L} the measurements support a scenario where HD~173701 has solar-like differential rotation, i.e., fastest at the equator and declining towards higher latitudes, where active regions have modulated the photometric time series. For comparison, the Sun has a rotation period ranging from 24.5 days at the equator to 28.5 days for the active regions at the highest latitudes \citep{1996ApJ...466..384D}. 

The rotation profile recovered from the peak-height ratios is also consistent with a solar-like differential rotation, with a relative differential rotation of $\sim0.53$ which is more than three times stronger than the solar value of $\sim 0.15$ \citep{Snodgrass1983,Snodgrass1990,1996ApJ...466..384D}.

Differential rotation is one of the hardest parameters to measure in other stars. Here we have used three different methods (activity modulation of the photometry, asteroseismology and peak-height ratios) and together they all provide a consistent picture of HD~173701 as a star with strong solar-like differential rotation. It is however true that the significance of any of the individual methods is low. It is also possible that we obtain different values with the different analysis methods because they are all wrong. However, we find this possibility unlikely given that all three methods have been tested on other stars and validated with solar observations.

\section{Discussions}
The asteroseismic analysis suggests that HD~173701 is truly Sun-like (except for the metallicity). The higher metallicity will increase the opacities inside the star, thereby lowering the luminosity. This will result in a slightly lower temperature and smaller radius. Based on the comparison between the temporal variability of the radial, dipolar and quarupolar oscillation modes, it also suggests that the dynamo driving the variability in HD~173701 is similar to the dynamo driving the solar cycle. The variability of both the chromospheric emission and especially the relative flux is however, significantly larger. We identify two possible explanations for this larger variability. Either the higher metallicity simply leads to a stronger dynamo with resulting stronger differential rotation, or the higher metallicity and lower inclination leads to a higher contrast of the facular component.

\begin{figure}[!t]
\plotone{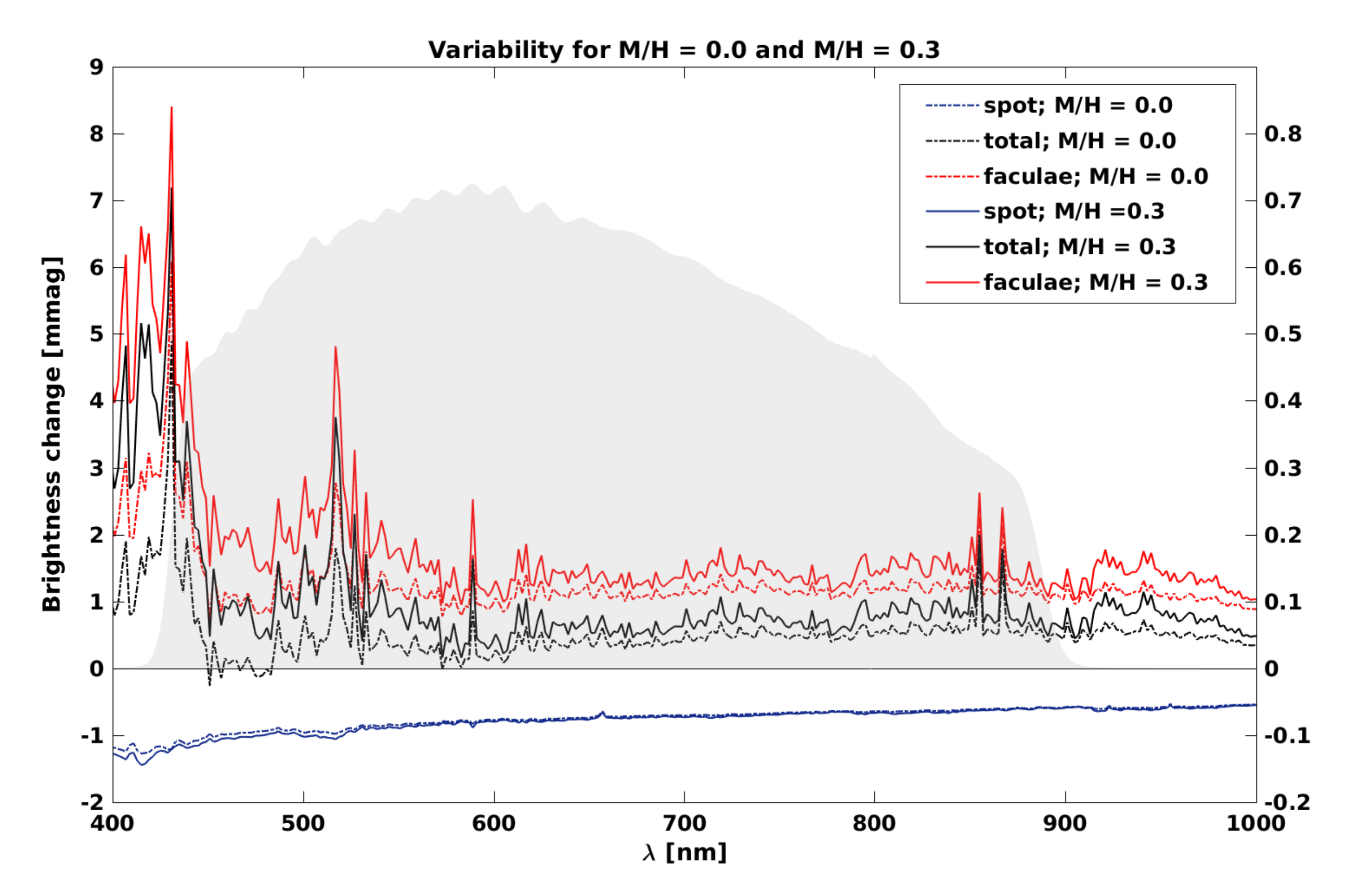}
\caption{Spectral irradiance variability as a function of metallicity. The plot shows solar photometric variability (dot-dashed lines) as well as that calculated for the hypothetical Sun (HD~173701) with 0.3 dex metallicity (M/H), which is measured as the logarithmic abundance of elements heavier than helium relative to the Sun (solid lines). Black, red, and blue lines are total, facular, and spot components of the variability, respectively. The shaded area indicates {\it Kepler} spectral efficiency. The variability of the relative photometry is found to be 1.4 higher in the {\it Kepler} bandpass for the high metallicity model compared to the Sun. If the contribution from the lower inclination of HD~173701 is included as well, the variability of the relative photometry is found to be 1.9 times higher in the {\it Kepler} bandpass compared to the Sun (see Fig. 9).}
\end{figure}

The facular contrast is strongly influenced by the Fraunhofer lines, which are affected by metallicity \citep{2015A&A...581A.116S}. Together with the inclination of the rotation axis, metallicity is also known to have an effect on the visibility of activity-related phenomena like spots and faculae \citep{2014A&A...569A..38S}. Unfortunately, this effect is poorly constrained because there are few metal-rich solar-analogs with measured activity cycles and inclinations. Activity cycles in Sun-like stars have so far mostly been discovered using observations from the so-called Mount Wilson HK Project \citep{1991ApJS...76..383D}, but most of these stars lack precisely determined stellar properties, especially ages, as there are no extensive asteroseismic observations of them. The main reason for this is that most of the stars in the Mount Wilson HK project were too bright to be observed by {\it Kepler}, which has so far been the main asteroseismic observatory. Moreover, {\it Kepler} only observed a limited part of the sky during its nominal mission \citep{2010ApJ...713L..79K}. 

The atmosphere of HD~173701 can affect the standard deviation of the relative photometry in two ways. The lower inclination leads to reduced visibility of both spots and faculae on the surface \citep{2014A&A...569A..38S}, and the facular contrast is strongly enhanced with metallicity through the effect of the weak atomic and molecular lines \citep{2015A&A...581A.116S}. Both effects would increase the standard deviation of the measured relative photometry of HD~173701. 

We have quantified the effect of metallicity and inclination on the standard deviation of the relative photometry followed the approach in \cite{2016A&A...589A..46S} and employed the SATIRE-S model \citep{2011JASTP..73..223K} to obtain solar brightness changes between 2000 and 2008 as they would have appeared if the Sun had been observed by $Kepler$ with different inclinations. We note that such a change roughly corresponds to the amplitude of solar cycle 23. SATIRE-S decomposes the solar disk into the quiet Sun and magnetic features (spot umbra, spot penumbra, and faculae) and sums up their contributions to return a time-dependent solar spectrum. For that the spectra of the quiet Sun and magnetic features at different disk positions are pre-calculated \citep{1999A&A...345..635U} by the ATLAS9 code \citep{1994A&A...281..817C}. To account for the high metallicity of HD 173701 we have recalculated these spectra for a M/H value of 0.3 dex using opacity distribution functions, which were synthesised with the DFSYNTHE code \citep{2005MSAIS...8...14K, 2005MSAIS...8...34C}. The opacity distribution functions were calculated from atmosphere models developed by \citet{1999A&A...345..635U}, where the effect of metallicity on the atmosphere's structure and electron concentration was neglected.

Fig.~8 demonstrates that the increase of the metallicity has only a subtle effect on the spot component of solar variability, whereas it significantly amplifies the facular component. Overall, the metallicity change from M/H=0.0 (solar value) to M/H=0.3 increases the amplitude of brightness variability as it would appear in {\it Kepler} observations from 0.48 milli magnitudes (corresponding to 0.044\%) to 0.84 milli magnitudes (0.077\%). Interestingly, changing the inclination from the solar value to $i=38^{\circ}$ has only a minor effect on the brightness variability of a hypothetical Sun with M/H=0.3 (Fig.~9), where an increase from 0.84 milli magnitudes (0.077\%) to 0.88 milli magnitudes (0.081\%) occurs.

\begin{figure}[!t]
\plotone{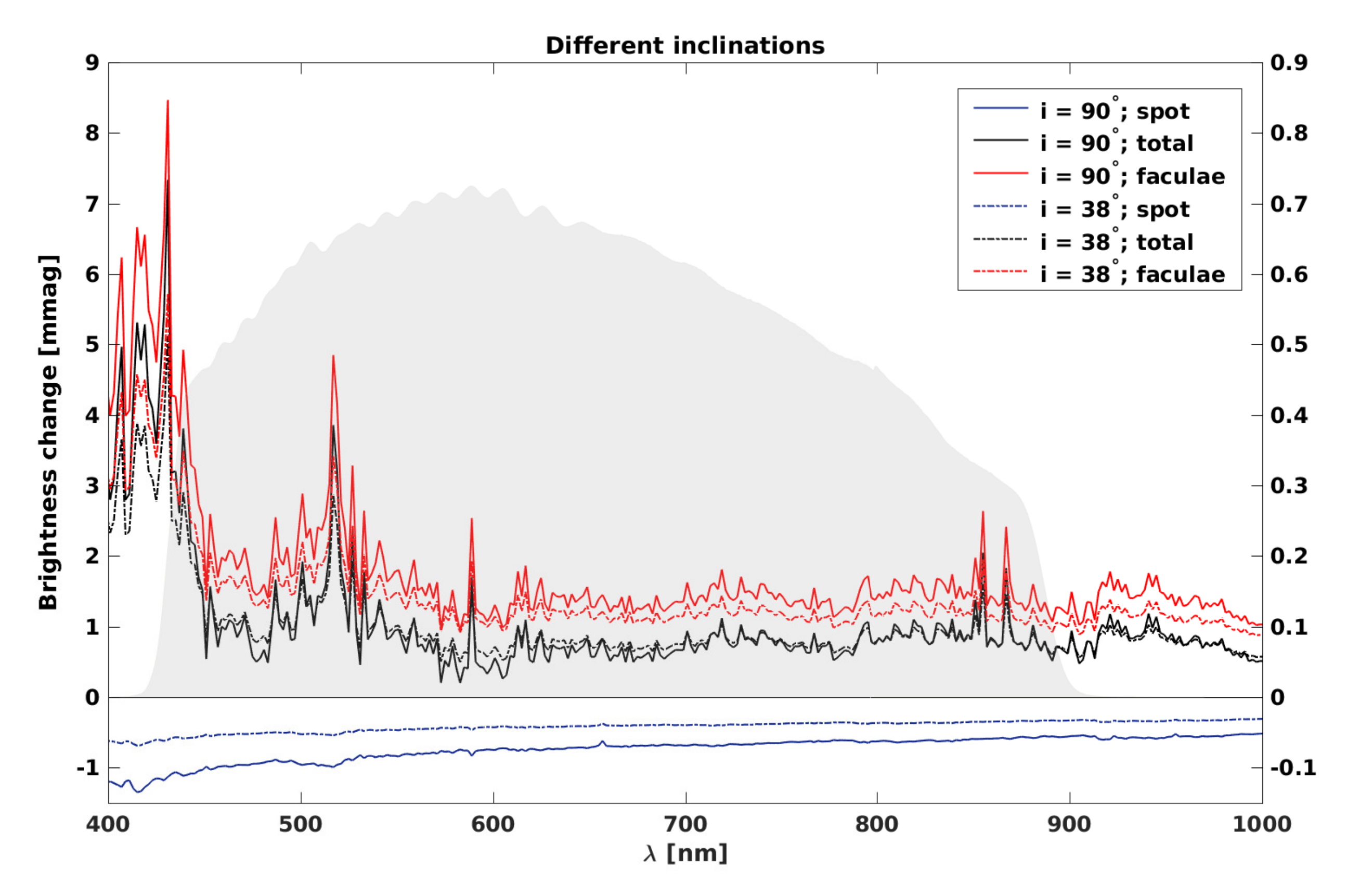}
\caption{Spectral irradiance variability as a function of inclination. The plot shows solar photometric variability (solid lines) as well as that calculated for the hypothetical Sun (HD~173701) with inclination of 38$^{\circ}$ (dot-dashed lines). Black, red, and blue lines are total, facular, and spot components of the variability, correspondingly. The shaded area indicates {\it Kepler} spectral efficiency. The variability of the relative photometry is found to be 1.4 times higher in the {\it Kepler} bandpass for the high metallicity mode compared to the Sun.}
\end{figure}

Increasing the metallicity of a Sun-like star will increase the opacities, which in turn will increase the temperature gradient. This means that the criterion for convection is satisfied deeper in the star \citep{1906WisGo.195...41S}. In this way, a doubling of the metallicity, as in HD~173701, leads to a convective zone that is approximately 8\% deeper than the solar convection zone \citep{2012ApJ...746...16V}. Theoretical studies have shown that a deeper convection zone leads to a longer convective turnover time near the base of the outer convection zone \citep{2017ApJ...836..192B} and thus stronger differential rotation \citep{2011ApJ...728..115B} in the same region. Stronger differential rotation will lead to a stronger dynamo (especially a stronger $\Omega$-effect). It is however, difficult to estimate exactly how much stronger the dynamo and the resulting cycle will be and also whether and how the stronger differential rotation will migrate all the way up to the surface. What we observe is strong surface differential rotation. We do not know if the radial differential rotation of HD~173701, which is likely what is important for the dynamo, is different from the Sun. 

We suggest that the most likely cause of the higher variability in the relative photometry and the chromospheric emission and the strong differential rotation of HD~173701 is the higher metallicity of the star. In this picture, the higher variability in the chromospheric emission and the strong differential rotation in HD~173701 is caused by a stronger dynamo induced by the higher metallicity. The large ratio between the variability seen in the relative flux and the chromospheric emission is caused by the higher facular contrast resulting from the combined effect of metallicity and inclination. 

One problem with this picture is however, that we only have good spectroscopic, photometric and seismic observations of the Sun over the last few decades. It is possible that the variability in the spectroscopic, photometric and seismic parameters of the Sun were significantly different during the Dalton or Maunder minimum. Though it is still a matter of debate, what caused the Dalton and Maunder minimum \citep{2010LRSP....7....3C} metallicity is out of the question as a possible cause.  In other words, though metallicity seems like the most likely cause of the stronger variability we observe in HD~173701 over the course of its 7.4-year activity cycle, we cannot rule out all other causes. 

The fact that the variability we observe in HD~173701 is much stronger then what we observe in the Sun, but the mean rotation periods are very similar, suggests that whatever is causing the stronger dynamo, the mean rotation period cannot be the driver. Thus under the simplest assumption that the behavior of the Sun during the last few decades is typical for the Sun, our suggestion that the most likely cause of the higher variability in the relative photometry and the chromospheric emission and the strong differential rotation in HD~173701 is the higher metallicity of the star is very plausible. We do however, urge observers to undertake a similar analysis on any solar twin with higher metallicity than HD~173701 in order to test this hypothesis.

\acknowledgments
We would like to thank the referee for thoughtful comments, which significantly improved the paper. The project is been supported by the Villum Foundation. Funding for the Stellar Astrophysics Centre is provided by the Danish National Research Foundation (Grant agreement No.: DNRF106). This work was partially supported by the Non-profit Adopt a Star program administered by White Dwarf Research Corporation. ARGS acknowledges the support from NASA Grant NNX17AF27G. ARGS, PPV, MSC and TLC acknowledges the support by the fellowship SFRH/BD/88032/2012 funded by FCT (Portugal) and POPH/FSE (EC), and from the University of Birmingham. Work by B.T.M. was performed under contract with the Jet Propulsion Laboratory (JPL) funded by NASA through the Sagan Fellowship Program executed by the NASA Exoplanet Science Institute. M.N.L. acknowledges the support of The Danish Council for Independent Research -- Natural Science (Grant DFF-4181-00415). SM acknowledges support from NASA grant NNX15AF13G and NSF grant AST-1411685 and the Ramon y Cajal fellowship number RYC-2015-17697. G.R.D and W.J.C. acknowledge the support of the UK Science and Technology Facilities Council (STFC). MSC acknowledges support from FCT through the research grant UID/FIS/04434/2013 and the InvestigadorFCT Contract No. IF/00894/2012 and the POPH/FSE (EC) by FEDER funding through the Programa Operacional de Factores de Competitividade (COMPETE). VW, AIS and NAK acknowledge the support by the European Research Council (ERC) under the European Union\'s Horizon 2020 research and innovation programme (grant agreement No. 715947) and by the German Federal Ministry of Education and Research under project 01LG1209A. The Nordic Optical Telescope is operated by the Nordic Optical Telescope Scientific Association at the Observatorio del Roque de los Muchachos, La Palma, Spain, of the Instituto de Astrof{\' i}sica de Canarias. The HK$\_$Project$\_$v1995$\_$NSO data derive from the Mount Wilson Observatory HK Project, which was supported by both public and private funds through the Carnegie Observatories, the Mount Wilson Institute, and the Harvard-Smithsonian Center for Astrophysics starting in 1966 and continuing for over 36 years.

\end{document}